\documentclass[12pt]{article}
\usepackage{amssymb} 
\usepackage{epsfig}
\usepackage{float}
\newcommand{\be}{\begin{equation}}
\newcommand{\ee}{\end{equation}}
\newcommand{\bea}{\begin{eqnarray}}
\newcommand{\eea}{\end{eqnarray}}

\begin{document} 

\begin{center}
{\bf A LECTURE ON NEUTRINOS }
\footnote{A talk at the colloquium at the University of Torino in December 2003.}
\end{center}

\begin{center}
S. M. Bilenky 
\end{center}
\vspace{0.1cm} 
\begin{center}
{\em  Joint Institute
for Nuclear Research, Dubna, R-141980, Russia\\}
\end{center}
\begin{center}
{\em 
SISSA, via Beirut 2-4, Trieste, 34014, Italy\\}
\end{center}
      
\begin{abstract}
The major steps in the development of our knowledge about neutrinos are 
reviewed.
The basics of neutrino oscillation formalism 
is presented. Neutrino oscillations in the framework of three-neutrino mixing
are considered. The evidence for neutrino oscillations 
is  
discussed.
\end{abstract}
\section{Introduction}
Neutrinos played a key role in the establishment of the
unified theory of the electromagnetic and weak interactions, the Standard Model (SM).
The first evidence for the unification (the discovery of  
Neutral Currents) was obtained in the neutrino experiments at CERN in 1973.
The measurement of the number of the flavor neutrinos in the LEP experiments in the nineties allowed to determine the number of quark-lepton families etc.

After the SM was fully established in high precision LEP and other experiments,
the most important discovery  in elementary particle physics  was 
the recent evidence found
in the Super-Kamiokande \cite{S-K}, SNO \cite{SNO}, KamLAND \cite{Kamland} 
and other solar \cite{Cl,GALLEX,SAGE,S-Ksol} and 
atmospheric \cite{Soudan,MACRO} neutrino experiments {\em of a new phenomenon, neutrino oscillations.}

Observation of neutrino oscillations means that flavor lepton numbers $L_{e}$, 
$L_{\mu}$ and $L_{\tau}$ are not conserved by a neutrino mass term of the Lagrangian,
which is the source of neutrino masses and mixing.
Small neutrino masses and neutrino mixing  are
apparently generated by {\em a new, beyond the SM mechanism}.
Many further investigations 
must be performed, however, in order to reveal the true nature of the 
discovered phenomenon.

The history of neutrinos is very interesting. One of its striking feature
is that the most fundamental and pioneer ideas  
were only partially correct.
Great courage and great intuition was required 
to propose them.
\section{Some historical remarks}

Pauli assumed the existence of neutrino in 1930 in order to explain the continuous
$\beta$- spectrum.
He suggested the existence of 
a neutral, spin
1/2 particle with mass of the same order of magnitude as the electron mass
and penetration length much larger than that of the photon 
(Pauli called this new particle neutron). Pauli assumed that in
the $\beta$-decay of nuclei 
a neutrino is emitted together with the electron
$$ (A,Z) \to (A,Z+1) + e^- + \nu. $$
The released energy will be shared in this case by the electron and
the neutrino (which can not be detected in the  $\beta$-decay experiments)
and a continues spectrum of electrons is observed.
In spite of the fact this is the only way to explain 
continuous $\beta$-spectra in the framework of
the conservation of the total energy and momentum, {\em in the thirties
the very idea of a new particle was very courageous }.\footnote{ 
It is well known that Dirac in the 
paper ``A theory of electrons and protons'' \cite{Dirac1},
in which the idea of 
the electron sea  was presented for the first time, 
identified the holes in the sea with
protons. Later he explained that ``the
whole climate of opinion at that time was against new particles''(see \cite{WBook}) . 
In 1931
Dirac came to the conclusion that the hole must be a new particle with positive charge and a mass equal to the electron mass \cite{Dirac3}. Dirac predicted the positron before it was discovered by Anderson in 1932 \cite{Anderson} }

Pauli presented his idea of neutrino in the famous letter addressed 
to the participants of the Tubingen conference.
In the same letter Pauli assumed, however,  that neutrinos are constituents of nuclei. 

At that time nuclei were 
considered as bound states of protons and electrons.
This assumption was in a contraction with the theorem on the connection between 
spin and statistics. For example, $^{7} \rm{N}_{14}$ nucleus according to the 
electron-proton model 
must have half-integer spin (14 protons and 7 electrons).  From the experimental data it was known, however, that 
$^{7}\rm{ N}_{14}$ nuclei satisfy Bose-Einstein statistics.
Pauli thought that additional spin 1/2 neutral particles in nuclei will be
a remedy.
After the discovery of the neutron by J. Chadwick in 1932 \cite{Chadwick} the electron-proton model and the problem of spin of $^{7} N_{14}$ disappeared. 

Fermi was the first who understood that 
the electron-neutrino pair is {\em produced in the transition}:
\footnote{It took about one year for Fermi to realize how 
electron and neutrino (which are not constituent of nuclei) are produced. Even though
quantum field theory existed 
the idea of 
creation of new particles in transitions of known ones, was not so evident 
at that time as now. For Fermi analogy with emission of photons in 
electromagnetic transitions was very important (P.Pontecorvo, private communication).}

\begin{equation}
n \to p + e^- +\bar\nu.
\label{001}
\end{equation}

The effective Hamiltonian of the decay (\ref{001}), proposed by Fermi, had the form 
\begin{equation}
{\cal{H}}_{I}^{\beta}=
G_F \,\bar {p} \gamma_{\alpha} n \,~\bar{e} \gamma^{\alpha} \nu + \rm{h.c.},
\label{002}
\end{equation}
where $G_F$ is the interaction constant (Fermi constant).
By analogy with electromagnetic interaction
\begin{equation}
{\cal{H}}_I^{\rm{em}} = 
e\, \bar{p} \gamma_{\alpha} p\,
A^{\alpha}
\label{003}
\end{equation}
Fermi assumed that 
$\beta$-decay interaction is a vector interaction.

The Hamiltonian (\ref{002}) describes allowed $\beta$-decay transitions in which
spins and parities of the initial and final nuclei are the same (Fermi selection rule):
$$ \Delta I =0\,~~ \pi_{i}= \pi_{f}.$$
This Hamiltonian cannot describe, however, 
decays,  which satisfy Gamov-Teller selection rule:

$$ \Delta I =\pm 1, 0\,~~ \pi_{i}= \pi_{f}.$$

Thus, it was clear in the thirties that the analogy between $\beta$-decay 
and electromagnetic interactions can be only 
partially correct.

The most general four-fermion
Hamiltonian of the $\beta$-decay 
was proposed by Gamov and Teller in 1936. It
has the form of the sum of the products of scalars (S), vectors (V),
tensors (T), axial vectors (A) and pseudoscalars (P):

\begin{equation}
{\cal{H}}_{I}^{\beta}=
\sum_{i=S,V,T,A,P}G_{i}\, \bar{p}\, O_{i}\, n \,~\bar{e}\, O^{i}\,  \nu  + 
\rm{h.c}.
\label{004}
\end{equation}
Here
$$O_{i} \to  1,\, \gamma_{\alpha},\,\sigma_{\alpha\,\beta},\,
\gamma_{\alpha}\gamma_{5},\, \gamma_{5}\,.$$ 
and $G_{i}$ are constants.

For many years the main aim of $\beta$-decay experiments
 was the search for the dominant terms in the Hamiltonian 
(\ref{004}). Up to 1956 the situation with the Hamiltonian of the 
$\beta$-decay was unclear and contradictory.

In 1956 the violation of parity in the $\beta$-decay was discovered in the Wu et al. \cite{Wu} 
and other experiments. In the Wu et al. experiment $\beta$-decay of the polarized 
$^{60}\rm{Co}$
was studied. 
In the general case of the non conservation of parity the probability of
the emission of the electron in the $\beta$ -decay of a nucleus with the polarization $\vec{P}$ is given by the following expression
\begin{equation}
w_{\vec P}(\vec k)=w_{0} \,(1 +\alpha \vec P \cdot \vec k )=
w_{0} \,(1 +\alpha P \,\cos \theta ),
\label{005}
\end{equation}
where $\vec k = \vec p/p $ is the unit vector in the direction of the momentum of the 
electron. If parity is conserved
$$ w_{\vec P}(\vec k)=w_{\vec P}(-\vec k)$$
and $\alpha =0$ (the pseudoscalar $\vec P \cdot \vec k$ can enter into the expression for the decay probability if the parity is conserved). In the Wu et al experiment it was found that
$\alpha \simeq - 0.7$ . Thus, large violation of parity in the $\beta$-decay 
was discovered in the experiment.

The violation of parity in the $\beta$-decay means that the
Hamiltonian of the process is a sum of scalar and pseudoscalar. The first 
Hamiltonian of this type was proposed by Lee and Yang in 1956 \cite{LeeYang1} before the Wu et al. experiment has been done.
The Hamiltonian, they considered, was a direct 
generalization of the Fermi-Gamov-Teller Hamiltonian (\ref{004}) and had the form
\begin{equation}
{\cal{H}}_{I}^{\beta}=
\sum_{i=S,V,T,A,P} \bar{p}\, O_{i}\, n \,~\bar{e}\, O^{i}\,(G_{i}
-G_{i}^{'}\gamma _{5})\,  \nu  + 
\rm{h.c.},
\label{006}
\end{equation}
where the constants $G_{i}^{'}$ characterize the pseudoscalar part of the Hamiltonian.
In the  Hamiltonian (\ref{006}) ten (!) arbitrary constants enter. 

The major progress in the establishment of the true effective Hamiltonian of the 
weak interaction was reached in 1957-58.
Two decisive steps were done at that time. {\em The first step was connected 
with neutrino}.

In 1957 Landau \cite{Landau}, Lee and Yang \cite{LeeYang2} and Salam 
\cite{Salam} proposed {\em the theory of the two-component 
neutrino}. Pauli considered the neutrino a particle with a mass 
different from zero and of the order of the electron mass $m_{e}$. At the time of the discovery of the parity violation 
from the experiments on the measurement of the high-energy part of the $\beta$- spectrum of
$^{3} H$,  it was found that neutrino mass $m$  is much less than the mass of the electron:
$$m \lesssim 200\, \rm{eV}\ll m_{e}.$$

The authors of the two-component neutrino theory assumed that {\em neutrino is massless particle}.
For massless
neutrino exists a very attractive possibility for large violation 
of the parity in the processes of emission (and absorption) of neutrino.

The field of neutrino $\nu(x)$
in the general case $m \neq 0$ satisfies the Dirac equation
\be
(i\, \gamma^{\alpha}\, \partial _{\alpha} -m)\,  \nu(x) =0.
\label{007}
\ee
For the left-handed and right-handed components
\be
\nu_{L,R}(x) = \frac{1\mp \gamma_{5}}{2}\nu(x)\,,
\label{008}
\ee
from (\ref{007}) we have {\em the system of two coupled equations}

\be
i\, \gamma^{\alpha}\, \partial _{\alpha}\,\nu_{L} -m\,\nu_{R} =0;\,~
i\, \gamma^{\alpha}\, \partial _{\alpha}\,\nu_{R} -m\,\nu_{L} =0.
\label{009}
\ee

If $m=0$ from (\ref{009}) we obtain  {\em the decoupled Weyl equations}: 
\be
i\, \gamma^{\alpha}\, \partial _{\alpha}\,\nu_{L,R}(x) =0.
\label{010}
\ee
Thus, in the case of the massless neutrino for neutrino 
field $\nu_{L}(x)$ (or $\nu_{R}(x)$) 
can be chosen.
\footnote{The equations (\ref {010}) for $\nu_{L}(x)$ and $ \nu_{R}(x)$ were
discussed by Pauli in his famous encyclopedia article ``General Principles of Quantum Mechanics'' \cite{Pauli}.  Pauli stressed that these equations are not invariant under space 
reflections and ``therefore are not applicable to the physical reality.''}
This was the choice, made by Landau, Lee and Yang and Salam.

If the neutrino field is $\nu_{L}(x)$ ($\nu_{R}(x)$ ) :
\begin{enumerate} 
\item

Parity is strongly violated :

\be
G'_{i}= G_{i} \,~~(G'_{i}= -G_{i} ).
\label{011}
\ee

\item
The helicity of neutrino is equal to 1 (-1) and helicity of antineutrino is equal to
-1 (1). 
\end{enumerate} 

The helicity of neutrino was measured in the famous Goldhaber et al. experiment in
1958 \cite{Goldhaber}. In this experiment the circular polarization of $\gamma$-quanta produced in the process
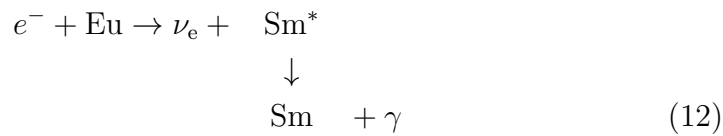
\begin{eqnarray}
e^- + \rm {Eu} \to \nu_e + \null & \rm {Sm}^* &
\nonumber
\\
& \downarrow &
\nonumber 
\\
& \rm{Sm} & \null + \gamma
\label{012}
\end{eqnarray}
was measured. The measurement of the polarization of the $\gamma$-quanta allows to 
determine the helicity of neutrino. It was found that 
\be
h = -1 \pm 0.3
\label{013}
\ee
Thus, the result of the Goldhaber et al. experiment 
was in agreement with the two-component neutrino theory. It was proved that of the
two possibilities $\nu_{L}(x)\,~ \rm{or} \,~ \nu_{R}(x)$, the first one is realized.
\footnote{We know today that neutrino masses are different from zero. 
However, due to the smallness of neutrino masses, two-component theory is an extremely good approximation.}

The next decisive step in the creation of the effective Hamiltonian of the weak interaction was
done by Feynman and Gell-Mann \cite{FG}, Marshak and Sudarshan 
\cite{Sudarshan-Marshak}. These authors 
assumed that not only neutrino field but {\em all fermion fields } enter in the 
Hamiltonian of the weak interaction in the form of the left-handed components.
\footnote {Let us notice that this was a courageous proposal: 
at the time when it  was done there were some $\beta$ -decay 
experiments, which were in contradiction with assumption made in  
\cite{FG,Sudarshan-Marshak}.
Later it occurred that the experiments were wrong.}

If only left-handed components of the fermion fields enter into the Hamiltonian,
the  most general four-fermion Hamiltonian of the $\beta$-decay is 
characterized by only one
constant and has the form
\be
{\cal{H}}_{I}^{\beta}=
\frac {G_{F}}{\sqrt{2}}4\, \bar{p}_{L}\gamma_{\alpha}  n_{L} \,~
\bar{e}_{L}\gamma^{\alpha}  \nu_{L}  + \rm{h.c.} 
\label{014}
\ee
The theory with the Hamiltonian (\ref{014})
was very successful: it allowed to describe the data of all  $\beta$-decay 
experiments.

Feynman and Gell-Mann introduced {\em the very important notion of charged 
weak current}
\be
j^{\alpha}=2 \ [\bar{p}_L \gamma^{\alpha} n_L \ + \ 
\bar{\nu}_{eL} \gamma ^{\alpha} e_L \ + \
\bar{\nu}_{\mu L} \gamma^{\alpha} \mu_L].
\label{015}
\ee
and assumed that the Hamiltonian of the weak interaction has the 
current $\times$ current,  $V-A$
form
\begin{equation}
{\cal{H}}_I =
\frac{G_{F}}{\sqrt{2}} \ j^{\alpha} \
j^{+}_{\alpha}\,.
\label{016}
\end{equation}

This Hamiltonian allowed to describe not only $\beta$ -decay but also $\mu$-capture, $\mu$-decay 
and other weak processes.

In the Feynman and Gell-Mann paper \cite{FG},  a possible origin of the current $\times$ current 
interaction was discussed. They considered the weak interaction theory with the intermediate heavy vector 
$W^{\pm}$ boson. If we assume that the true Lagrangian of the weak interaction has the form
\begin{equation}
{\cal{L}}=-\frac{g}{2\sqrt{2}} j_{\alpha} W^{\alpha} \ + \rm{h.c.},
\label{017}
\end{equation}
where $g$ is the dimensionless interaction constant, the effective Hamiltonian of the processes with the virtual $W$ boson with momentum $Q$ , which satisfies the condition 
$Q^2\ll m^2_{W}$,  has the current $\times$ current form (\ref{016}). The Fermi 
constant is connected 
with the constant $g$ by the relation
\begin{equation}
\frac{G_{F}}{\sqrt{2}} = \frac{g^{2}}{8\,m_{W}^{2}}\,.
\label{018}
\end{equation}

With the further development of the theory of the weak interaction the notion of the charged weak current drastically changed. In 1962 in the Brookhaven experiment \cite{Danby}
it was shown that electron neutrino $\nu_{e}$ and muon neutrino $\nu_{\mu }$
are different particles.\footnote{ The idea of this experiment was proposed by B. Pontecorvo in 1959 \cite{BPontecorvo59}}

In 1962 strange particles were included into the weak charged current by 
N. Cabibbo \cite{Cabibbo}. Later with the establishment of the notion of quarks, fields of nucleons in the weak current were replaced by the fields of quarks.
In 1970 after the Glashow-Illiopulous-Maiani paper \cite{GIM}, 
 the charged weak current took the form
\begin{equation}
j_{\alpha} = 2\,~[\bar{\nu}_{eL} \gamma_{\alpha} e_L +
\bar{\nu}_{\mu L} \gamma_{\alpha} \mu_L +
\bar{u}_L \gamma_{\alpha} d'_L +
 \bar{c}_L \gamma_{\alpha} s'_L]\,,
\label{019}
\end{equation}
where $c(x)$ is the field of the {\em predicted} charmed quark and $d'_L$ and  $s'_L$ are orthogonal combinations (mixture) of the  $d_L$ and  
$s_L$ fields
\begin{eqnarray}
d'_L & = & \cos{\theta_C}\, d_L + \sin{\theta_C}\, s_L
\nonumber \\
s'_L & = & -\sin{\theta_C}\,d_L + \cos{\theta_C}\, s_L\,,
\label{020}
\end{eqnarray}
where $\theta_{C}$ is the Cabibbo angle.

First charmed particles were successfully 
discovered in 1975. The data on the investigation of the decays of strange and charmed particles
and neutrino processes were in  good agreement with the mixing relations (\ref{020}).

After the establishment of the quark mixing (\ref{020}) 
it was natural to assume that quark-lepton analogy 
holds: neutrino masses like masses of all other 
fundamental fermions (quarks and leptons) are different from zero and fields of massive neutrinos 
enter into charged current  in a mixed form
\begin{eqnarray}
\nu_{e L}& = & \cos{\theta}\, \nu_{1L}+ \sin{\theta}\, \nu_{2L}
\nonumber \\
\nu_{\mu L} & = & -\sin{\theta}\,\nu_{1L} + \cos{\theta}\, \nu_{2L}\,,
\label{021}
\end{eqnarray}
where $\theta$ is a lepton mixing angle.
Neutrino oscillations and other consequences of the neutrino mixing (\ref{021}) were considered in the seventies
(see \cite{BP78}).

Up to now we have discussed the phenomenological period of the 
development of the theory of the weak interaction. 
The real theory, the Standard Model, was created in the seventies (Glashow \cite{Glashow61}, 
Weinberg \cite{Weinberg67}, Salam \cite{Salam68}).
This theory was born in an attempt
to construct a renormalizable theory of the weak interaction. The only physical 
renormalizable theory, known at that time, was quantum electrodynamics. 
The success in the creation
 of the renormalizable theory of the weak interaction was reached 
on the way of the {\em unification} of the weak and electromagnetic interactions into one electroweak interaction. 

The SM is based on $SU(2)\times U(1)$ local gauge invariance and the Higgs mechanism of the spontaneous violation of symmetry.  The SM summarizes the development of elementary particle physics in the last century and is one of the greatest achievement 
in physics. 
The SM Lagrangian of the interaction of fermions and vector bosons has the form
of the sum of the charged current (CC), neutral current (NC) and electromagnetic terms:
\begin{equation}
{\cal{L}}_I = (-\frac{g}{2\sqrt{2}} j^{CC}_{\alpha} W^{\alpha} + 
h.c.) - \frac{g}{2 \cos{\theta_W}} j^{NC}_{\alpha} Z^{\alpha} -
e j_{\alpha}^{\rm{em}} A^{\alpha}.
\label{022}
\end{equation}
 Here $j^{CC}_{\alpha}$,  $j^{NC}_{\alpha}$ and  $j_{\alpha}^{\rm{em}} $ are charged, neutral and electromagnetic currents, $ W^{\alpha}$ and $ Z^{\alpha}$ are fields of $ W^{\pm}$ and 
$ Z^{0}$ vector bosons, $ A^{\alpha}$ is the electromagnetic field  and 
$\theta_{W}$ is the weak angle (parameter of the theory).
 
The SM {\em predicted} the existence of $ W^{\pm}$ and $ Z^{0}$ vector bosons and their masses,
{\em predicted} new NC interaction,  {\em predicted} charmed, bottom, top and other particles etc. 
All prediction of the  SM are in perfect agreement with the data 
of many experiments, including very precise LEP  experiments. \footnote{
In spite of this impressive agreement with the existing data, the SM, as it is well known, can not be considered as
a satisfactory theory . There are several open problems in the SM. They are mainly 
connected with the Higgs mechanism of the spontaneous
violation of the symmetry: hierarchy problem,  the problem of the large number of 
free parameters (fermion masses, mixing angles, CP-phases ) etc.}

The SM {\em provides a natural framework for the quark mixing }
After spontaneous violation of symmetry for the quark charged current we have  

\begin{equation}
j_\alpha^{CC} = 2 [
\overline{u}_L \gamma_\alpha d'_L
+ \overline{c}_L \gamma_\alpha s'_L
+\overline{t}_L \gamma_\alpha b'_L ], 
\label{023}
\end{equation}

where

\begin{equation}
d'_L = \sum_{q=d,s,b} V_{uq} q_L;\,~ s'_L = \sum_{q=d,s,b} V_{cq} q_L ;\,~
b'_L = \sum_{q=d,s,b} V_{tq} q_L.
\label{024}
\end{equation}
Here $V$ is the unitary Cabibbo-Kobayashi-Maskawa (CKM) quark mixing matrix, which is characterized by three mixing angles and one CP phase \cite{Cabibbo, K-M}.

In 1967 when Weinberg and Salam papers appeared only two charged leptons $e$ and $\mu$
and two neutrinos $\nu_{e}$ and $\nu_{\mu}$ were known. After the success of the theory 
of the two-component neutrino there was at that time general opinion
that neutrinos were massless particles. The original 
SM was build for two charged leptons and {\em two massless neutrinos}.

Neutrino masses and mixing can be introduced, however, in the framework of the SM with the Higgs doublet in the same manner as masses and mixing of quarks are introduced.
Neutrino fields $\nu_{lL}$ ($l=e, \mu, \tau$) that enter in this case into the lepton charged current
\begin{equation}
j_\alpha^{CC} = 2 [
\overline{\nu}_{eL} \gamma_\alpha e_L
+ \overline{\nu}_{\mu L} \gamma_\alpha \mu _L
+\overline{\nu}_{\tau L}  \gamma_\alpha \tau _L], 
\label{025}
\end{equation}
are mixtures of the fields of neutrinos with definite masses
\begin{equation}
\nu_{{l}L} = \sum_{i=1}^{3} U_{{l}i} \, \nu_{iL} \,
\label{026}
\end{equation}
where $\nu_{i}$ is the field of neutrino with the 
mass $m_{i}$ and $U$ is the unitary mixing matrix. Like the CKM matrix the matrix $U$
is characterized by three mixing angles and one CP phase.

We will come back to the relation (\ref{026}) later. 
Now we will discuss the {\em birth of the idea of neutrino mixing and oscillations}. 
This idea was proposed by B. Pontecorvo in 1957-58 \cite{P57,P58} soon after the two-component 
neutrino theory appeared. At that time only the electron neutrino was known.

B. Pontecorvo was fascinated by the $K^0 \rightleftharpoons  \overline{K^0}$
oscillations and looked for the analogous phenomenon in the lepton world.
For the first time he mentioned a possibility of transitions of
$\nu  \rightleftharpoons \bar \nu$  in vacuum in 1957 in
the paper \cite{P57}, in which the transitions of muonium into antimuonium 
($(\mu^{+}e^{-})\rightleftharpoons(\mu^{-}e^{+})$) were discussed. 
 In this paper he wrote
\begin{quote}
``If the two-component neutrino theory turn out to be incorrect
(which at present seems to be rather improbable)
and if the conservation law of neutrino charge would not apply,
then in principle neutrino $\rightleftharpoons $ antineutrino transitions
could take place in vacuum.''
\end{quote}

In the fifties Reines and Cowan have been doing their famous experiment \cite{ReinesCowan} 
in which $\bar \nu_{e}$'s 
from a reactor were discovered via the observation of $e^{+}$ and  $n$ in 
the process 
$$\bar \nu_{e} +p \to e^{+} +n.$$
In 1957-58 another experiment with reactor antineutrinos was done by R. 
Davis \cite{Davis}. In this
experiment the process
\be
\bar \nu_{e} +^{37}\rm{Cl} \to e^{-} +^{37}\rm{Ar},
\label{027}
\ee
in which the lepton number is violated, was searched for.
 A rumor reached  B. Pontecorvo that R. Davis observed 
events (\ref {027}). 
He came to the conclusion that neutrino oscillations could be a possible and natural explanation of the ``observed events''
and published the  first paper on neutrino oscillations \cite{P58}.

B. Pontecorvo assumed that neutrino and antineutrino, produced in usual weak processes, are
different particles and there exists an additional interaction, which transfer neutrino into antineutrino.
He concluded that in this case 
``neutrino and antineutrino are {\em mixed} particles, i.e., a symmetric and
antisymmetric combination of two truly neutral Majorana particles $\nu_1$
and $\nu_2$ of different combined parity'' .\footnote {In the first papers 
on neutrino oscillations B.Pontecorvo considered maximum mixing}.''

There were two parts in the Pontecorvo paper \cite{P58}. First
he suggested that in the Reines and Cowan experiment

\begin{quote}
``The cross section of the
production of neutrons and positrons in the process of the absorption of antineutrinos from a reactor by protons would be smaller than the expected cross section....It would be 
extremely interesting to perform  the Reines and Cowan experiment at different distances from reactor.''
\end{quote}
In 2002 the effect, predicted by B. Pontecorvo in 1958, was observed in the 
reactor KamLAND experiment \cite{Kamland}, in which the  average distance between reactors and the detector 
is about 180 km. Of course,  B. Pontecorvo could not known the distance at which the effect can be observed.  He noticed in  \cite{P58}
\begin{quote}
``Effects of transformation of neutrino into antineutrino and vice versa may be unobservable in the laboratory  because of the large values of R, but will certainly occur on an astronomical scale.''
\end{quote}

In the second part of the paper \cite{P58} B. Pontecorvo tried to explain ``events'' (\ref{027}).  He wrote 
\begin{quote}
``It is not possible to state a priori that antineutrino flux,  which at the beginning can not 
initiate the reaction (\ref{027}), is not transferred into a flux some part of which  could produce
this reaction''
\end{quote}

In the framework of the two-component theory right-handed
antineutrinos, produced in decays of neutron-reach nuclei in a reactor, 
 can not induce the reaction 
(\ref{027}).
Later, when the two-component neutrino theory was established
 B. Pontecorvo understood that {\em in the case of one neutrino} right-handed antineutrino
$\bar\nu_{e}$ can transfer {\em only} into right-handed neutrino $\nu_{eR}$, which is {\em a sterile particle}.
In fact, B. Pontecorvo was the first who introduced \cite{P67} the notion of sterile neutrinos  so popular nowadays.

After the second neutrino $\nu_{\mu}$ was discovered in the Brookhaven experiment it was not difficult for 
B. Pontecorvo to generalize his idea of neutrino oscillations
for the case of two types of neutrinos \cite{P67} .
In 1967 before the first results of the R. Davis solar neutrino experiment were published he 
{\em envisaged the solar neutrino problem:}. 

\begin{quote}
``From an observational point of view the ideal object is the sun. If
the oscillation length is smaller than the radius of the sun region
effectively producing neutrinos, (let us say one tenth of the sun radius
$R_\odot$ or 0.1 million km for $^8B$ neutrinos, which will give the
main contribution in the experiments being planned now), direct
oscillations will be smeared out and unobservable. The only effect on
the earth's surface would be that the flux of observable sun neutrinos
must be two times smaller than the total (active and sterile) neutrino
flux.''
\end{quote}

In 2002-03 this prediction was confirmed in a model-independent way 
by the SNO solar neutrino experiment \cite{SNO}.

With these remarks we conclude the part connected with the history of neutrino
 and we  come 
to the discussion of the present status of neutrino oscillations.

\section{Neutrino oscillations}

In all solar \cite{SNO,Cl,GALLEX,SAGE,S-Ksol} and 
 atmospheric \cite{S-K,Soudan,MACRO} and in the reactor KamLAND  \cite{Kamland} 
neutrino experiments 
{\em compelling evidence of neutrino oscillations was obtained}.
The interpretation of the data of these experiments is based on the following assumptions:
\begin{enumerate}
\item
The Lagrangians of the interaction of neutrinos with other particles are the standard 
CC and NC Lagragians (\ref{022}) with the lepton charged current and neutrino neutral current 
 given by the expressions
\begin{equation}
j^{\mathrm{CC}}_{\alpha} = \sum_{l=e,\mu,\tau} \bar \nu_{lL} \gamma_{\alpha}l_{L};\,~~
j^{\mathrm{NC}}_{\alpha} =\sum_{l=e,\mu,\tau} \bar \nu_{lL}\gamma_{\alpha}\nu_{lL}\,.
\label{028}
\end{equation}
\item
Tree flavor neutrinos $\nu_{e}$, $\nu_{\mu}$ ,$\nu_{\tau}$ (and antineutrinos
$\bar \nu_{e}$, $\bar \nu_{\mu}$ ,$\bar \nu_{\tau}$ ) 
exist in nature.
\item
Neutrino mixing takes place:
\begin{equation}
\nu_{{l}L} = \sum_{i}U_{{l}i} \, \nu_{iL} \,,
\label{029}
\end{equation}
where
$\nu_{i}$ is the field of neutrino  with the 
mass $m_{i}$ and $U$ is the unitary mixing matrix.
The field $\nu_{i}$ can be the Dirac field if the total lepton number $L$ is conserved.
If there are no conserved lepton numbers, $\nu_{i}$ is the field of the Majorana neutrino, which
satisfies the condition
$$ \nu_{i}=\nu^{c}_{i}=C \bar \nu^{T}_{i},$$
where $C$ is the matrix of the charge conjugation.

\end{enumerate}

Neutrino oscillations is a new phenomenon. There are many 
discussions of the quantum mechanical problems connected with
it (see, for example, \cite{Giunti}and references therein ). We will present here the field-theoretical point of view 
(see \cite{BP78,BPet,BilGiunti}),
which allow us to obtain the standard formulas for transition probabilities and to understand the origin of oscillations (neutrino, neutral kaons etc).

Flavor neutrinos $\nu_{e}$, $\nu_{\mu}$  and $\nu_{\tau}$ are produced in decays of pions, 
kaons, neutron-reach nuclei in a reactor, neutrino reactions in the sun etc. Let us consider the emission of neutrinos
with momentum $\vec p$  in a decay
\be
a \to b + l^{+}+ \nu_{l}\,.
\label{030}
\ee
In the case of the neutrino mixing (\ref{029}) 
the state of the final particles is given by
\be
 |f\rangle = \,~~\sum_{i=1}^{3} U^{*}_{li}\, |\nu_i\rangle \, | l^{+}\rangle \,|b\rangle 
 \,~ \langle \nu_i\,l^{+}\,b\,| S |\,a\rangle \,.
\label{031}
\ee
Here $\langle \nu_i\,l^{+}\,b\,| S |\,a\rangle $ is the transition matrix element 
 and $|\nu_i\rangle$ is the state of neutrino with the momentum
$\vec{p}$ and the energy $E_{i} =\sqrt{p^2 + m^{2} _{i}}$.

From data of the Mainz \cite{Mainz} and the Troitsk \cite{Troitsk} 
tritium experiments the following upper bound on the absolute value of 
neutrino mass
was obtained
$$m_{i} \leq 2.2\,\rm{eV}$$
A more strict bound 
$$m_{i} \leq 0.6 \,\rm{eV}$$
was found from the analysis of the latest cosmological data \cite{Sloan}.

Energies $E$ of neutrinos in neutrino experiments are in the MeV range (solar and reactor experiments)
in GeV range (atmospheric, accelerator long baseline experiments) etc.
Thus,   $(m_{i}/E)^2\ll 1$ and we can safely neglect tiny effects of neutrino masses
in production processes.
For the decay (\ref {030})
we have
\be
\langle \nu_i\,l^{+}\,b\,| S |\,a\rangle\simeq \langle \nu_i\,l^{+}\,b\,| S |\,a\rangle_{m^{2}_i=0}
=\langle \nu_l\,l^{+}\,b\,| S |\,a\rangle_{SM} ,
\label{032}
\ee
where $\langle \nu_l\,l^{+}\,b\,| S |\,a\rangle_{SM}$ is the SM matrix element of the process
(\ref{030}).
For the state $|f\rangle$ we have 
\be
 |f\rangle \simeq    |\nu_l\rangle \, |l^{+}\rangle \,|b\rangle 
 \,\langle \nu_l\,l^{+}\,b\,| S |\,a\rangle_{SM}  \,,
\label{033}
\ee
where
\be
|\nu_l\rangle =\sum_{i=1}^{3} U^{*}_{li}\,~ |\nu_{i}\rangle
\label{034}
\ee
is the state of the flavor neutrino $\nu_{l}$.
\footnote{In the experiments on the direct measurement of the neutrino mass \cite{Mainz,Troitsk} 
the high energy
part of the spectrum of electrons from the decay of tritium is studied. 
This part of the spectrum corresponds to the emission of neutrinos with small energies.
Effect of the distortion of the electron spectrum can be
observed if neutrino energy and neutrino mass are of the same order.
The relation (\ref{032}) obviously is not valid in this case.} 

The relation (\ref{034}) is similar to the relations
\bea
|K^{0}\rangle = \frac{1}{2N}\,  (|K_{S}^{0}\rangle+ |K_{L}^{0}\rangle)
\nonumber\\
|\overline K^{0}\rangle = \frac{1}{2N}\, \frac{p}{q}\,(|K_{S}^{0}\rangle- |K_{L}^{0}\rangle),
\label{035}
\eea
which connect states of the neutral kaons $K^{0}$ and $\overline K^{0}$, particles with definite strangeness, 
with  the states  of $K_{S}^{0}$ and $ K_{L}^{0}$, particles with 
definite masses and widths.
In (\ref{035}) $N= \frac{1}{\sqrt{1+ |\frac{q}{p}|^{2}}}$ is the normalizing factor and 
$p/q$ is a complex parameter. 

The relation (\ref{035}) is based on the Weisskopf-Wigner 
approximation, which is valid in the case of neutral kaons because the mass difference
$\Delta m = m_{K_{L}}- m_{K_{S}}$ is much smaller than the kaon mass. 

The states of flavor neutrinos (\ref{034}) are 
{\em coherent superpositions} of the states
of the neutrino $\nu_{i}$ with definite mass.
For the flavor antineutrino $\bar\nu_{l}$ we have
\be
|\bar\nu_l\rangle =\sum_{i=1}^{3}U_{li}\,~ |\bar\nu_{i}\rangle,
\label{036}
\ee
where $|\bar\nu_{i}\rangle$ is the state of antineutrino (in the Dirac case) or  neutrino (in the Majorana case) with mass $m_{i}$, momentum $p=(\vec p, E_{i})$ 
and positive helicity ( $|\nu_i\rangle$
in Eq. (\ref{034}) is the state with negative helicity).

We will consider now the evolution  of the flavor states (\ref{034}) in vacuum.
If at $t=0$ flavor neutrino $\nu_{l}$ is produced, for the neutrino state at a time $t$ we
will have
\be
|\nu_{l}\rangle_{t}=\,~ e^{-iH_{0}\,t}\,~|\nu_{l}\rangle =
\sum_{1}^{2}\,U_{l i}^*\,e^{-iE_{i}t}\,|\nu_{i}\rangle,
\label{037}
\ee
where $H_{0}$ is the free Hamiltonian.
Developing $E_{i}$ over $m_{i}^2$ we have
\be
E_{i}\simeq E + \frac{m^2_{i}}{2\,E},
\label{038}
\ee
where $E=p$  is the energy of the neutrino in the approximation $m^2_{i}\to 0$.
From (\ref{037})  and (\ref{038}) for the neutrino state at the time $t$  we have
\be
|\nu_{l}\rangle_{t}=\,~ e^{-iE\,t}\,~
\sum_{i=1}^{3} e^{-i\,\frac{m^2_{i}\,t}{2E}} \,~U^{*}_{li}\,|\nu_{i}\rangle.
\label{039}
\ee
Thus, at  a time $t$ {\em different massive neutrino states $|\nu_{i}\rangle$ acquire
different phases.}
The phase differences at macroscopic distances $L\simeq t$
can be large. This means that the flavor content of the state $|\nu_{l}\rangle_{t}$ 
can be quite
different from  the initial state $|\nu_{l}\rangle$. The differences of the phases of the different
neutrino mass components at macroscopic distances is {\em the physical origin of the phenomenon of neutrino oscillations.} 

Let us discuss now a detection
process. 
Neutrinos are detected via the observation of CC and NC weak processes. 
Taking into account the unitarity of the mixing matrix, from 
(\ref{034})  we have 
\be
|\nu_{i}\rangle =\sum_{l}U_{li}\,~ |\nu_l\rangle.
\label{040}
\ee
From  (\ref{039}) and (\ref{040}) we find
\be
|\nu_{l}\rangle_{t}=\,~ e^{-iE\,t}\,~
\sum_{l'} \,{\mathrm A}(\nu_{l} \to \nu_{l'}) \,~|\nu_{l'}\rangle,
\label{041}
\ee
where
\be
{\mathrm A}(\nu_{l} \to \nu_{l'}) =\sum_{i=1}^{3}U_{l'i}\,~ e^{-i\,\frac{m^2_{i}\,t}{2E}} \,~U^{*}_{li}
\label{042}
\ee
is the amplitude of the probability to find a
state $|\nu_{l'}\rangle$ in the state
$|\nu_{l}\rangle_{t}$.

Let us consider 
the CC process
\be
\nu_{l'}+ N\to l' +X.
\label{043}
\ee 
If the neutrino state is 
$|\nu_l\rangle $,
the matrix element of the production of the lepton $l'$ in the process (\ref{043}) is given by
\be
\langle  l' X| S |\nu_{l} N\rangle = \sum_{i}\langle  l' X| S |\nu_{i} N\rangle\,~U^{*}_{li}\,~ U_{l'i}.
\label{044}
\ee
Now, taking into account that $(m_{i}/E)^2\ll 1$, we have

\be
\langle  l' X| S |\nu_{i} N\rangle \simeq \langle  l' X| S |\nu_{i} N\rangle _{m^2_{i}=0}
=\langle  l' X| S |\nu_{l'} N\rangle _{SM},
\label{045}
\ee
where $\langle  l' X| S |\nu_{l'} N\rangle _{SM}$ is the SM matrix element of the process
(\ref{043}).

Using now the unitarity of the mixing matrix, from (\ref{044}) and 
(\ref{045}) we obtain
\be
\langle  l' X| S |\nu_{l} N\rangle \simeq \delta_{l l'}\,~\langle  l' X| S |\nu_{l'} N\rangle _{SM}.
\label{046}
\ee

From  (\ref{041}), (\ref{042}) and  (\ref{046}) for the probability 
of the transition of the flavor neutrino $\nu_{l}$,  produced in a standard CC weak process, to the flavor neutrino 
$\nu_{l'}$ , detected in a standard CC weak process, we obtain the following expression
\begin{equation}
{\mathrm P}(\nu_{l} \to \nu_{l'}) =|\,\sum_{i=1}^{3} U_{l' i} 
\,~ e^{- i \Delta m^2_{i1} \frac {L} {2E}} \,~U_{l i}^*\, |^2,
\label{047}
\end{equation}
where $\Delta m^2_{i1} =  m^2_{i}- m^2_{1}$ \footnote{ We label neutrino masses in such a way 
that $m_1<m_2<m_3$}  $L$ is the distance between neutrino source and neutrino 
detector ($L\simeq t$ , where $t$ is the time between production and detection of 
neutrinos\footnote{The relation $L=t$ was used in the long baseline accelerator K2K experiment\cite{K2K} in order to provide timing information for the selection of neutrino events in the Super-Kamiokande detector.} )

Taking into account the unitarity of the mixing matrix we can present 
the transition probability in the form 
\begin{equation}
{\mathrm P}(\nu_{l} \to \nu_{l'})= |\,\delta_{l' l} +\sum_{i=2,3} U_{l' i} 
\,~ (e^{- i \Delta m^2_{i1} \frac {L} {2E}} -1) U_{l i}^*\, |^2.
\label{048}
\end{equation}

The following remarks are in order.
\begin{enumerate}
\item
 For the probability of the antineutrino transition  $\bar\nu_{l} \to \bar\nu_{l'}$ 
we have
\begin{equation}
{\mathrm P}(\bar\nu_{l} \to \bar\nu_{l'}) =
|\,\delta_{l' l} +\sum_{i=2,3} U^* _{l' i} 
\,~ (e^{- i \Delta m^2_{i1} \frac {L} {2E}} -1) U_{l i}\,|^2
\label{049}
\end{equation}
\item
From the comparison of  Eq.(\ref{048}) and Eq.(\ref{049}) we conclude that the following 
relation  
\be
{\mathrm P}(\nu_{l} \to \nu_{l'})= {\mathrm P}(\bar\nu_{l'} \to \bar\nu_{l}) 
\label{050}
\ee
holds. The relation (\ref{050}) is a consequence of the CPT theorem.
\item
In the case of the CP invariance in the lepton sector the mixing matrix
$U$ is real in the Dirac case. In the Majorana case 
the mixing matrix 
satisfies the condition 
\be
U_{\alpha i}=U_{\alpha i}^{*}\,~\eta_{i}\,,
\label{051}
\ee
where $\eta_{i}=\pm i$ is the CP parity of the Majorana neutrino $\nu_{i}$.
From (\ref{048}), (\ref{049}) and (\ref{051})
it follows that in the case of the CP invariance in the lepton sector
the probabilities of the transitions $\nu_l \to \nu_{l'} $ and $\bar\nu_l\to \bar\nu_{l'}$
are equal:
\be
{\mathrm P}(\nu_l \to \nu_{l'}) =
{\mathrm P}(\bar\nu_l\to \bar\nu_{l'})
\label{052}
\ee

\item
If the number of the neutrinos $\nu_{i}$ is larger than three, for the mixing we have
\be
\nu_{{l}L} = \sum_{i=1}^{3+n_{s}}U_{{l}i} \, \nu_{iL}; \,~~
\nu_{{s}L} = \sum_{i=1}^{3+n_{s}}U_{{s}i} \, \nu_{iL}, 
\label{053}
\ee
where $U$ is the unitary $(3+n_{s})\times (3+n_{s})$ mixing matrix and 
$\nu_{{s}L}$ ($s=s_{1},....s_{n_{s}}$) is a sterile field (the field which does not enter into CC and NC 
Lagrangians). In the case of small masses $m_{i}$ the states of the sterile neutrinos 
are {\em defined} as follows
\be
|\nu_s\rangle =\sum_{i=1}^{3+n_{s}} U^{*}_{si}\,~ |\nu_{i}\rangle
\label{054}
\ee
From the unitarity of the mixing matrix for the state of neutrino with definite mass we have
 in this case
\be
|\nu_{i}\rangle =\sum_{\alpha}U_{\alpha i}\,~ |\nu_{\alpha}\rangle,
\label{055}
\ee
where index $\alpha$ takes the values $e, \mu,\tau, s_{1},...$

The transition probability in the general case of the transitions into active (flavor) and sterile states is given by the expression 
\begin{equation}
{\mathrm P}(\nu_{\alpha} \to \nu_{\alpha'})= |\,\delta_{\alpha' \alpha} +\sum_{i= 2}^{3+n_{s}} U_{\alpha'  i} 
\,~ (e^{- i \Delta m^2_{i1} \frac {L} {2E}} -1) U_{\alpha i}^*\, |^2.
\label{056}
\end{equation}

\item
If neutrinos are detected via the observation of NC processes, the observed number of the NC 
events is the product of the total probability of the transition of the initial neutrino 
$\nu_{l}$
into all flavor states
$\sum _{l'=e,\mu,\tau} {\mathrm P}(\nu_{l} \to \nu_{l'})$ 
and the number of the events expected in case of no neutrino oscillations.
If there are no transitions into sterile states, from  
the conservation of probability we have 
$$\sum _{l'=e,\mu,\tau} {\mathrm P}(\nu_{l} \to \nu_{l'})=1.$$
Thus, there are no neutrino oscillations in this case. 

If there are transitions into sterile states, we have
$$\sum _{l'=e,\mu,\tau} {\mathrm P}(\nu_{l} \to \nu_{l'})=1- 
\sum _{s}{\mathrm P}(\nu_{l} \to \nu_{s}),$$
where $\sum _{s}{\mathrm P}(\nu_{l} \to \nu_{s})$ is the total probability of the transition of the 
flavor neutrino $\nu_{l}$ into all possible sterile states.

\item
Neutrino mass term does not conserve flavor lepton numbers $L_{e}$, $L_{\mu}$,
and $L_{\tau}$. What are flavor neutrinos $\nu_{e}$, $\nu_{\mu}$ and $\nu_{\tau}$?
It is clear from the previous discussion that flavor neutrino $\nu_{l}$
is a particle which is emitted together $l^{+}$ in
CC production processes,  produces $l^{-}$ in CC detection processes etc.
Neutrino oscillations take place because {\em the states of flavor neutrinos are
not states with definite masses}.

\item
In the simplest case of the transition between two flavor neutrinos index $i$ in 
(\ref{048}) and (\ref{049}) takes the value 2. For $l'\not=l$ we have
\begin{equation}
{\mathrm P}(\nu_{l} \to \nu_{l'})
=\frac {1}{2}\,~  \sin^{2}  2\theta \,~
 (1 - \cos \Delta m^{2} \frac {L} {2E});\,~~(l' \not= l).
\label{057}
\end{equation}
Here $\Delta m^{2} = m_{2}^{2}- m_{1}^{2}$ and $\theta$  is the mixing angle 
($|U_{l'2}|^2 = 
\sin^{2}\theta, \,~|U_{l2}|^2 =\cos^{2}\theta$).

It is easy to see that in the two-neutrino case the following relations are automatically satisfied
\be
{\mathrm P}(\nu_{l} \to \nu_{l'}) ={\mathrm P}(\nu_{l'} \to \nu_{l})={\mathrm P}(\bar\nu_{l} \to \bar\nu_{l'}).
\label{058}
\ee
Thus, the CP violation in the lepton sector can not be revealed if there are transitions only between
two neutrinos.

For the $\nu_{l}$ ($\nu_{l'}$) survival probability from the conservation of probability we find
\be
{\mathrm P}(\nu_l \to \nu_{l}) =
1-{\mathrm P}(\nu_l \to \nu_{l'});~~{\mathrm P}(\nu_{l'} \to \nu_{l'}) =
1-{\mathrm P}(\nu_{l'} \to \nu_{l})
\label{059}
\ee
Thus, we have

\be
{\mathrm P}(\nu_l \to \nu_{l}) ={\mathrm P}(\nu_{l'} \to \nu_{l'})
\label{060}
\ee
Let us stress that this relation is valid only in the case of the transitions between two types of neutrinos.

The formulas (\ref{057}) and (\ref{059}) describe periodical transitions 
$\nu_l \rightleftarrows \nu_{l'}$ with the oscillation length (in the units $h=c=1$) 
given by the expression
\be
L_{0}=4\,\pi \frac {E} {\Delta m^{2} } 
\label{061}
\ee
The oscillation length can be also presented in the form
\be
L_{0}= 2.48\frac {E} {\Delta m^{2}}\,  m,
\label{062}
\ee
where $E$ is neutrino energy in MeV and $\Delta m^{2}$ is neutrino mass-squared difference in
$\rm{eV}^{2}$.

From (\ref{062})   it follows that   in order to observe 
neutrino oscillations the condition 
$$L \gtrsim\ L_{0};\,~~\rm{or} \,~~\Delta m^{2}\gtrsim \frac{E}/{L}$$
must be satisfied.

\end{enumerate}

\section{Evidence for neutrino oscillations}
In the solar neutrino experiments Homestake \cite{Cl}, 
GALLEX-GNO \cite{GALLEX}, SAGE \cite{SAGE} and  Super-Kamiokande \cite{S-Ksol},  neutrinos are detected in 
different energy ranges. In all these experiments the observed event rates are significantly 
smaller that the rates predicted by the Standard Solar Model \cite{SSM}. 
This suppression can be 
naturally explained by the transition of the solar $\nu_{e}$'s into  
other neutrino neutrinos.

Impressive evidence of such transition was obtained recently in the SNO experiment \cite{SNO}.
In this experiment high energy solar neutrinos, produced in the decay 
$^8 \rm{B} \to ^8 \rm{Be} +e^{+} + \nu_{e}$ , were detected via the observation of the CC
process
\be
\nu_{e}+d\to e^{-} +p+p
\label{063}
\ee
and NC process

\be
\nu_{l}+d\to \nu_{l}+n+p\,~~ (l =e, \mu,\tau)
\label{064}
\ee

The measurement of the CC event rate allows to determine the total flux of the solar 
$\nu_{e}$ 's on the earth. In the SNO experiment it was found

\be
\Phi_{\nu_{e}}^{\rm{SNO}}=({1.59^{+0.09}_{-0.07}\mbox{(stat.)}^{+0.06}_{-0.08}~\mbox{(syst.)}} ) \cdot 10^{6}\,~
cm^{-2}s^{-1} \,.
\label{065}
\ee

The measurement of the NC event rate allows to determine {\em the total flux 
of all high energy flavor neutrinos $\nu_{e}$,  $\nu_{\mu}$ and 
$\nu_{\tau}$ on the earth.} In the SNO experiment it was obtained
\be
\sum_{l=e,\mu,\tau}\Phi_{\nu_{l}}^{\rm{SNO}}=(5.21\pm 0.27\pm 0.38) \cdot 10^{6}\,~
cm^{-2}s^{-1} \,.
\label{066}
\ee

The SNO resuls 
clearly demonstrate that original solar  $\nu_{e}$'s
on the way from the production region in the central zone of the sun to the earth are
transferred into $\nu_{\mu}$ and $\nu_{\tau}$ .

The data of all solar neutrino experiments can be described 
by the two-neutrino MSW \cite{MSW} transitions in matter. From the analysis of
the solar neutrino data 
in the preferable LMA region the following  best-fit values 
of the two neutrino oscillation parameters
were found \cite{SNO}:
\be
\Delta m^{2}_{\rm{sol}}=5\cdot 10^{-5}\rm{eV}^{2};
\,\tan^{2}\theta_{\rm{sol}}=0.34;\,~
(\chi^{2}_{\rm{min}}= 57/72\, \rm{d.o.f.}).
\label{067}
\ee
Further compelling evidence in favor  of neutrino oscillations 
was obtained recently in the reactor KamLAND experiment \cite{Kamland}. For the value of 
$\Delta m^{2}_{\rm{sol}}$, given by  (\ref{067}), the oscillation length is in the range
$L_{0}\simeq (100-200) \,km$. 
In the  KamLAND experiment  $\bar\nu_{e}$'s  
from many reactors in Japan and Korea are 
detected via the observation of $e^{+}$ and $n$, produced in the process
$$\bar\nu_{e}+p \to e^{+}+n .$$
The threshold of this process is equal to 1.8 MeV.
About 80\% of the total 
number of the events is due to 
antineutrinos from 26 reactors 
within the distances 138-214 km.
For the ratio of total numbers of the observed and expected events 
the following value was found 
\be
\frac{N_{obs}}{N_{exp}} = 0.611 \pm 0.085 \pm 0.041.
\label{068}
\ee
From the analysis of all solar and  KamLAND data the following best-fit values of
neutrino oscillation parameters were found \cite{SNO}:
\be
\Delta m^{2}_{\rm{sol}}=7.1\cdot 10^{-5}\rm{eV}^{2};
\,\tan^{2}\theta_{\rm{sol}}=0.41;\,~.
\label{069}
\ee

Another important evidence in favor of neutrino oscillations was obtained in the atmospheric neutrino experiments \cite{S-K,Soudan, MACRO}. Atmospheric neutrinos are 
produced mainly in the decays of pions (produced in the interaction of the cosmic rays with nuclei in the earth atmosphere)
 and subsequent decays of
muons :$ \pi \to \mu \,~\nu_{\mu};\,~~ \mu \to e \,~\nu_{\mu}\,~\nu_{e}$.
In the Super-Kamiokande (S-K) experiment \cite{S-K}
significant dependence on the azimuthal angle $\theta_{z}$ of the Multi-Gev 
muon events ($E_{\rm{vis}}\geq 1.3$ GeV) was observed. Neutrino coming to the detector from above ( $0.2\leq \cos \theta_{z}\leq 1$) travel the distances $L$ in the range
$20\, \rm{km}\lesssim L\lesssim 500 \,\rm{km}$ and neutrinos which enter into the detector from
below ( $-1\leq \cos \theta_{z}\leq -0.2$) travel the distances  in the range
$500\,\rm{km}\lesssim L\lesssim 13000 \,\rm{km}$.
For the ratio of the total number of the up-going $\nu_{\mu}$ to the total number of the
down-going $\nu_{\mu}$ in the Super-Kamiokande experiment the following value
\be
\left(\frac{U}{D}\right)_{\mu}= 0.54 \pm 0.04 \pm 0.01.
\label{070}
\ee
was found.

The observation of the large up-down asymmetry 
clearly demonstrates the dependence of the number of the muon neutrinos
on the distance they
travel from the production region in the atmosphere to the detector.

The S-K data \cite{S-K} and data of other atmospheric neutrino experiments
(SOUDAN 2 \cite{Soudan}, MACRO \cite{MACRO} ) are perfectly described, 
if we assume two-neutrino $\nu_{\mu}\rightleftarrows\nu_{\tau}$ 
oscillations.
From the analysis of the S-K data 
the following
best-fit values of the neutrino oscillation parameters 
\be
\Delta m^{2}_{atm}=2\cdot 10^{-3}\rm{ eV}^{2};\,~\sin^{2}2 \theta_{atm}=1.0 
\,~(\chi^{2}_{\rm{min}}= 170.8/ 170\,\rm{d.o.f.}) 
\label{071}
\ee
were found \cite{S-K}.

Atmospheric neutrino evidence in favor of neutrino oscillations have been confirmed by the
accelerator long baseline K2K experiment\cite{K2K}. In this experiment muon neutrinos with 
average energy 1.3 GeV, produced at the KEK facility, were detected by the Super-Kamiokande 
detector at the distance of about 250 km. The total number of the observed muon events was equal to 56. The expected number of the muon events was equal to $80.1 ^{+6.2}_{-5.4}$.

Thus, in the K2K experiment indications in favor of the
disappearance  of the accelerator $\nu_{\mu}$ were obtained.
From the two-neutrino analysis of the data  the following 
best-fit
values of the oscillation parameters were found
\be
\sin^{2}2\,\theta_{K2K}=1:\,~ \Delta m^{2}_{K2K}= 2.8\,~10^{-3}\,~ \rm{eV}^{2}.
\label{072}
\ee
These values are compatible  with the values of the oscillation parameters that were found from the analysis of the S-K atmospheric neutrino data. 

\section{ Neutrino oscillations in the framework of three-neutrino mixing}

We will consider here neutrino oscillations in the framework of the three-neutrino mixing
\be
\nu_{lL} =\sum_{i=1}^{3} U_{li}\,\nu_{iL},
\label{073}
\ee
where $U$ is 3$\times$3   Pontecorvo-Maki-Nakagawa-Sakata (PMNS) \cite{P58,P67,MNS}  mixing matrix.

From the analysis of the data of neutrino oscillation experiments it was established that two 
independent neutrino mass-squared differences $\Delta m^{2}_{sol}$ and $\Delta m^{2}_{atm}$
satisfy the hierarchy

\be
\Delta m^{2}_{sol}\ll \Delta m^{2}_{atm}.
\label{074}
\ee
In the framework of the three-neutrino mixing 
neutrino oscillation data are compatible with two types of
neutrino mass spectra:

I. ``Normal'' mass spectrum 
\be
\Delta m^{2}_{21}\simeq \Delta m^{2}_{\rm{sol}};
\,
\Delta m^{2}_{32}\simeq \Delta m^{2}_{\rm{atm}};
\label{075}
\ee

II. ``Inverted'' mass spectrum \footnote{Notice that 
for inverted spectrum neutrino masses are often labeled in such a way that
$m_{3}<m_{1}<m_{2}$. In this case 
$$\Delta m^{2}_{21}\simeq \Delta m^{2}_{\rm{sol}};\,~|\Delta m^{2}_{31}|\simeq \Delta m^{2}_{\rm{atm}}$$}

\be
\Delta m^{2}_{32}\simeq \Delta m^{2}_{\rm{sol}};
\,
\Delta m^{2}_{21}\simeq \Delta m^{2}_{\rm{atm}};
\label{076}
\ee
Let us consider
neutrino oscillations 
in the atmospheric range of $L/E$ ($L/E \simeq 10^3$) under the assumption of the normal neutrino mass spectrum.
In this range the ``solar''  phase $\Delta m^{2}_{21}\,L/2E $ is small and
we can neglect the
$i=2$ term in Eq. (\ref{048}) 
and Eq. (\ref{049}) .
For the $\bar\nu_{e}$  survival  probability we have 
\be
{\mathrm P}(\bar\nu_e \to \bar\nu_e)= 
1 - \frac {1} {2}
{\mathrm B}_{e e}\,~ 
(1 - \cos \Delta m^{2}_{32} \frac {L} {2E})\,,
\label{077}
\ee
where
\be
{\mathrm B}_{e e}=
4\,~|U_{e3}|^{2}
\,~(1 -|U_{e 3}|^{2}).
\label{078}
\ee
is the amplitude of oscillations.

In two reactor experiments CHOOZ \cite{CHOOZ} and Palo Verde \cite{PVerde}
the search for neutrino oscillations in the atmospheric range of $L/E$ 
have been performed. The reactor-detector distances in these experiments were about 1 km. 
No disappearance of reactor $\bar \nu_e$'s was found. From the exclusion plot,
  obtained  from the analysis of the data 
of the CHOOZ experiment, at the point $\Delta m^{2}_{32} =2\cdot 10^{-3}\rm{eV}^{2}$ 
(the S-K best- fit value) for the amplitude ${\mathrm B}_{e e}$ 
we have the following upper bound 
\be
{\mathrm B}_{e e}\lesssim 2\cdot10^{-1}
\label{079}
\ee

From (\ref{078}) and (\ref{079}) it follows that the parameter $|U_{e 3}|^{2}$
can be small or large (close to 1). Taking into account the solar neutrino data
we can exclude the large values of $|U_{e 3}|^{2}$.
Thus, we have
\be
|U_{e 3}|^{2}=\sin^{2}\theta_{13}\lesssim5\cdot 10^{-2},
\label{080}
\ee
where $\theta_{13}$ is the 1-3 mixing angle.

Neglecting the solar term in (\ref{048}  )  and (\ref{049}), 
for the probability of $\nu_{\mu} \to  \nu_{\tau} $ ($\bar\nu_{\mu} \to  \bar\nu_{\tau}$) 
transition in the atmospheric range  
we find the following expression 

\begin{equation}
{\mathrm P}(\nu_\mu \to \nu_{\tau}) ={\mathrm P}(\bar\nu_\mu \to \bar\nu_{\tau})=
 \frac {1} {2} {\mathrm A}_{\tau \mu}\,~
\left (1 - \cos \Delta m^{2}_{32} \frac {L} {2E}\right),
\label{081}
\end{equation}
where  the oscillation amplitude is given by the expression
\be
{\mathrm A}_{\tau\mu}= 4\,~|U_{\tau 3}|^{2}\,~|U_{\mu 3}|^{2}.
\label{082}
\ee
In the standard parametrization of the neutrino mixing matrix we have
\be
U_{\mu 3}= \sqrt{1-|U_{e3}|^{2}} \sin\theta_{23};\,~U_{\tau 3}= \sqrt{1-|U_{e3}|^{2}} \cos\theta_{23},
\label{083}
\ee
where $\theta_{23}$ is the 2-3 mixing angle.
Thus, the amplitude of the $\nu_{\mu} \rightleftharpoons  \nu_{\tau} $ oscillations is given by

\be
A_{\tau;\mu}= (1-|U_{e3}|^{2})^{2}\sin^{2}2\,\theta_{23}
\label{084}
\ee

In the expression for
the $\bar\nu_e$ survival probability in vacuum in  the 
``KamLAND range'' of $L/E$
the effect of large  neutrino mass-
 squared difference  $\Delta m^{2}_{32}$  is averaged out.  We have

\be
{\mathrm P}(\bar\nu_{e} \to\bar\nu_{e})=
|U_{e 3}|^{4}+ (1-|U_{e 3}|^{2})^{2}\,~
P^{(1,2)}(\bar\nu_{e}\to\bar\nu_{e})\,,
\label{085}
\ee
where
\begin{equation}
{\mathrm P}^{(1,2)}(\bar\nu_e \to \bar\nu_e) =
 1 - \frac {1} {2}\,~\sin^{2}2\,\theta_{12}\,~ 
(1 - \cos \Delta m^{2}_{21} \frac {L} {2E})\,,
\label{086}
\end{equation}
is the two-neutrino transition probability in vacuum and $\theta_{12}$ is 1-2 mixing angle.

The $\nu_e$ survival probability in matter 
is given by the expression \cite{Schramm}
\be
{\mathrm P}_{\rm{mat}}(\nu_{e} \to \nu_{e})=
|U_{e 3}|^{4}+ (1-|U_{e 3}|^{2})^{2}\,~
P^{(1,2)}_{\rm{mat}}(\nu_{e}\to\nu_{e})\,,
\label{087}
\ee
where $P_{\rm{mat}}^{(1,2)}(\nu_{e}\to\nu_{e})$ is the two-neutrino survival probability in
matter, which depend on parameters $\Delta m^{2}_{21} $ and $\tan^{2}\theta_{12}$.

From the data of the CHOOZ experiment it follows that 
the parameter $|U_{e 3}|^{2}$ is small (see (\ref{080}) ).
If we will neglect the contribution of 
$|U_{e 3}|^{2}$ to the transition probabilities, 
the following simple picture of neutrino oscillations emerges (see \cite{BGiu}):
\begin{enumerate}
\item
In the atmospheric range of $L/E$ there are no transitions $\nu_{e} \to \nu_{\mu,\tau}$
and
neutrino oscillations in this range  
are pure {\em two-neutrino} $\nu_{\mu} \rightleftharpoons  \nu_{\tau} $ oscillations.
Taking into account (\ref{081}) and (\ref{084}),
we have
\be
\sin^{2}2\theta_{23}\simeq \sin^{2}2\,\theta_{atm};\,~ 
\Delta m^{2}_{32}\simeq \Delta m^{2}_{atm}.
\label{088}
\ee
\item

The solar $\nu_{e}$ are transfered inside of the sun into 
$\nu_{\mu}$ and $\nu_{\tau} $.
These transitions are described by 
the {\em two-neutrino survival probability in matter.}
We have 

\be
\tan^{2}\,\theta_{12}\simeq \tan^{2}\,\theta_{\rm{sol}}; \,~ 
\Delta m^{2}_{12}\simeq \Delta m^{2}_{\rm{sol}}.
\label{089}
\ee

\item
In the KamLAND range of $L/E$  $\nu_{e} \rightleftarrows \nu_{\mu, \tau}$ oscillations take place. 
The $\bar \nu_{e} $ survival probability is given in this range
by {\em the standard two-neutrino expression.} 
Thus we have
\be
\sin^{2}2\,\theta_{12}\simeq \sin^{2}2\,\theta_{K-L};\,~  
\Delta m^{2}_{21}\simeq \Delta m^{2}_{K-L}.
\label{090}
\ee
\item
In the standard parametrization
\be
U_{e 3} = \sin\theta_{13}\,~e^{-i\,\delta},
\label{091}
\ee
where $\delta$ is the CP phase. Thus, in the leading approximation 
the effects of the
CP violation in the lepton sector can not be revealed.

\end{enumerate}

This picture is in a good agreement with the existing neutrino oscillation data.
The next important step in the investigation of neutrino oscillations will be the search for possible 
small effects beyond the leading approximation.

First of all it is necessary to measure the value of the parameter $\sin^{2}\theta_{13}$
(or to improve the bound (\ref{080})).
Several groups are considering a possibility to perform a new reactor 
experiment of the CHOOZ type with a reactor-detector distance about 1 km 
(see Reactor White Paper \cite{13}).
Information about the parameter
$\sin^{2}\theta_{13}$ can be also obtained from experiments on the search for 
$\nu_{\mu} \to \nu_{e} $
oscillations in the future long baseline accelerator neutrino experiments. 
Such experiments will be done
by the MINOS collaboration \cite{MINOS} and under preparation at the JPARC facility 
\cite{JPARC}.

\section{Neutrinoless double $\beta$-decay}

The search for neutrinoless double $\beta$-decay ($0\nu \beta\beta$- decay) 
\be
(\rm{A,Z}) \to (\rm{A,Z +2}) +e^- +e^-.
\label{092}
\ee
is one of the most important problem of today's neutrino physics. The observation of this 
decay would be a proof that the total lepton number is not conserved and massive 
neutrinos $\nu_{i}$ are Majorana particles. The half-life of the process 
is given by (see \cite{Doi,BPet,Elliott}):
\be
\frac{1}{T^{0\,\nu}_{1/2}(A,Z)}=
|m_{ee}|^{2}\,|M^{0\,\nu}(A,Z)|^{2}\,G^{0\,\nu}(E_{0},Z)\,.
\label{093}
\ee
Here
\be
m_{ee}=\sum_{i}U_{ei}^{2}\, m_{i}
\label{094}
\ee
is the effective Majorana mass,
$|M^{0\,\nu}(A,Z)|$ is the nuclear matrix element
and $G^{0\,\nu}(E_{0},Z)$
is  known phase-space factor ($E_{0}$ is the energy release).
Many experiments on the search for $0\nu \beta\beta$ decay are going on at present (see \cite{Gratta}).
The best lower bound on the half-life was reached in the  $^{76} \rm{Ge}$ Heidelberg-Moscow experiment \cite{H-M}:
\be
T^{0\nu}_{1/2}(^{76} \rm{Ge})\geq 1.9 \cdot 10^{25}\, y\,~~ (90\% \,\rm{CL})
\label{095}
\ee
Taking into account the results of different calculations of the nuclear matrix element, 
from this bound for the effective Majorana mass the following upper bound can be inferred
\be
|m_{ee}| \lesssim (0.3-1.2)\,~\rm{eV}\,. 
\label{096}
\ee
Several new experiments on the search for the neutrinoless 
double $\beta$-decay of 
$^{76} \rm{Ge}$, $^{130} \rm{Te}$, $^{136} \rm{Xe}$,  $^{100} \rm{Mo}$ and other
nuclei are 
in preparation (see \cite{Gratta}).
In these experiments significant 
improvement in the sensitivity to $|m_{ee}|$
(one-two orders of the magnitude )  is expected.

The effective Majorana mass $|m_{ee}| $ are determined by the absolute value of neutrino masses and elements $U_{ei}^{2}$. For the neutrino masses we have

\be
m_{2} =\sqrt{m^{2}_{1}+\Delta m^{2}_{\rm{21}}};\,
m_{3} =\sqrt{m^{2}_{1}+\Delta m^{2}_{\rm{21}}+ \Delta m^{2}_{\rm{32}}}\,,
\label{097}
\ee
where parameters $\Delta m^{2}_{\rm{21}}$ and $\Delta m^{2}_{\rm{32}}$ can be determined from
the data of neutrino oscillation experiments. 
The elements $U_{e1}$ and $U_{e1}$ are given by
\be
U_{e1}=\sqrt{1-|U_{e3}|^{2}} \,\cos\theta_{12}\, e^{i\alpha_1}\,~
U_{e2}=\sqrt{1-|U_{e3}|^{2}} \,\sin \theta_{12}\, e^{i\alpha_1}\,,
\label{098}
\ee
where $\alpha_1$ and  $\alpha_2$ are Majorana CP phases. The value of the parameter 
$\sin^{2} \theta_{12}$ and upper bound of the parameter $|U_{e3}|^{2}$ 
can be determined 
from the neutrino oscillation data.

The possible value of the effective Majorana mass strongly depends on the minimal neutrino mass and character of neutrino mass spectrum (see \cite{BGGM,PP}).
In the case of neutrino mass hierarchy 

\be
m_{1} \ll m_{2} \ll m_{3}. 
\label{099}
\ee
neutrino masses $m_{2}$ and $m_{3}$ are given by 
\be
 m_{2}\simeq \sqrt{ \Delta
m^{2}_{\rm{sol}}};\,~~
m_{3}\simeq \sqrt{ \Delta m^{2}_{\rm{atm}}}
\label{100}
\ee
and minimal neutrino mass is small (
$m_{1}\ll 
\sqrt{ \Delta m^{2}_{\rm{sol}}} $).
For the effective Majorana mass we have
in this case the bound
\be
|m_{ee}|\lesssim 4.6\cdot 10^{-3}\,\rm{eV},
\label{101}
\ee
which is much smaller than the sensitivity of the future experiments on the search for
$0\nu \beta\beta$- decay.

In the case of the inverted mass hierarchy
\be
m_{1} \ll m_{2} < m_{3}. 
\label{102}
\ee
for neutrino masses we have
\be
m_{2} \simeq m_{3}\simeq
\sqrt{ \Delta m^{2}_{\rm{atm}}}
\label{103}
\ee
and $m_{1}\ll\sqrt{ \Delta m^{2}_{\rm{atm}}}$.

The effective Majorana mass $|m_{ee}|$ 
in this case is given by
\be
|m_{ee}|\simeq \sqrt{ \Delta m^{2}_{\rm{atm}}}\,~
(1-\sin^{2} 2\,\theta_{\rm{sol}}\,\sin^{2}\,\alpha)^{\frac{1}{2}}\,,
\label{104}
\ee
where $\alpha=\alpha_{2}-\alpha_{1}$ is Majorana CP-phase
difference.
From (\ref{104}) we obtain the range
\be
2.9\cdot 10^{-2}\, \rm{eV} \lesssim
 |m_{ee}|\lesssim 5.5\cdot 10^{-2}\, \rm{eV} .
\label{105}
\ee
The value of the effective Majorana mass in this range can be reached in the future
 $0\nu \beta\beta$- decay experiments. 

If the minimal neutrino mass is relatively large 
( $m_{1}\gg\sqrt{ \Delta m^{2}_{\rm{atm}}})$) in this case neutrino masses are practically degenerate.  For the effective 
Majorana mass we have in this case
\be
0.65 \,m_{1}\leq
 |m_{ee}|\leq m_{1}
\label{106}
\ee
Let us notice that in the future tritium experiment KATRIN  \cite{Katrin}  the sensitivity $m_{1}\simeq 0.25 \rm{eV}$
is planned to be reached.

\section{A remark on the see-saw mechanism}

The understanding of the physical origin of neutrino masses and mixing is the  main purpose
of modern experimental  and theoretical neutrino physics.
From existing data we know that neutrino masses are much smaller than masses 
of quarks and leptons. 
For example, for the masses of the fundamental fermions of the third family we have
\bea
m_{t} =174.3\pm 5.1\, \rm{GeV};\, m_{b}=(4.0-4.5)\rm{GeV}; \\
\nonumber
m_{\tau}=1776.99\pm 0.29 \, \rm{
MeV};\, m_{3} \leq 2.2 \rm{eV}\,~ (0.6 \rm{eV})
\label{107}
\eea
The original Standard Model  \cite{Glashow61,Weinberg67,Salam68} 
was build for the case of massless two-component neutrinos.
In the framework of the SM there is, however, no principle 
which force neutrino masses to be equal to zero (like gauge invariance for the photon).
In the SM masses of quarks and leptons are parameters, generated 
by the Higgs mechanism with one Higgs doublet. 
On the same footing masses of neutrinos can be generated.
We may expect, however, that neutrino masses, generated by the standard Higgs mechanism, 
 are of the same order of magnitude 
as masses of other  family partners (quarks and lepton). Thus, to be in agreement with experimental data
we need in this case an additional beyond the SM mechanism
which provide smallness of neutrino masses.

One of the most attractive mechanism of such type is the see-saw \cite{see-saw}.
Let us consider the simplest case of one generation. The Dirac mass term 

\be
\mathcal{L}^{\mathrm{D}} = -m\,\bar \nu_{R}\nu_{L} +\mathrm{h.c.} 
\label{108}
\ee
with $m$ of the order of a lepton or quark mass 
can be generated by the standard Higgs mechanism.

Let us assume that there exists a new mechanism which does not conserve lepton number 
and generate the right-handed Majorana mass term 
\footnote{The conservation of the electric charge does not allow such a term
for quarks and charged leptons}
\be
\mathcal{L}^{\mathrm{M}}_{R} = -\frac{1}{2}\,M\,\bar \nu_{R}(\nu_{R})^{c} +\mathrm{h.c.}, 
\label{109}
\ee
where $\nu_{R}$ is $SU(2)$ singlet, $(\nu_{R})^{c} = C\, \bar\nu^{T}_{R}$ is the charge 
conjugated field and
$M \gg m$ (usually it is assumed that $M \simeq M_{\rm{GUT}}\simeq
(10^{15}-10^{16})\, \rm{GeV}$.

After the diagonalization of the total mass term we have
\bea
\nu_{L} &=& i\,\cos \theta \,
\nu_{1L} +\sin \theta\, \nu_{2L}\nonumber\\ 
(\nu_{R})^{c} &=& -i\,\sin \theta \,\nu_{1L} +\cos \theta\, \nu_{2L}, 
\label{110}
\eea
where $\nu_{1}$  
is the field of the Majorana neutrino with the mass
\be
m_{1}\simeq \frac{m^{2}}{M}\ll m
\label{111}
\ee
and $\nu_{2}$ is the field of a heavy neutral Majorana lepton with the mass
\be
m_{2}\simeq M \gg m
\label{112}
\ee
The mixing angle $\theta$ is given by the relation
\be
\tan 2\,\theta = \frac{2\,m}{M}\ll 1
\label{113}
\ee
In the general case of three families the see-saw mechanism generates 
three light Majorana masses and three heavy Majorana masses .  
Let us stress that if  neutrino masses are of the standard see-saw origin :
\begin{itemize}
\item
Neutrinos with definite masses are Majorana particles.
\item
There are three light neutrinos.

\item

Three heavy Majorana leptons, see-saw partners of neutrinos, must exist.

\end{itemize}
The existence of the heavy Majorana particles,
could be a source of the baryon asymmetry of the Universe (see \cite{Buch}).

\section{Conclusion}

The history of neutrino demonstrates the complicated and unpredictable way of science. We may 
expect further surprises from neutrinos. 
The discovery of neutrino oscillations, driven by small neutrino masses and 
neutrino mixing, requires apparently either additional to SM mechanism of neutrino mass generation or a completely new mechanism, very different from the mechanism of 
generation of 
masses of quarks and leptons.
In order to reveal the true nature 
of the new phenomenon we definitely need additional information.

The great progress in neutrino physics, reached in the recent years,  dictates the nearest 
problems to be solved:

\begin{itemize}
\item
What is the nature of neutrinos with definite masses ?

Are they Dirac particles, possessing the conserved lepton number, or
truly neutral Majorana particles?
Investigation of neutrino oscillations do not allow to answer this question 
\cite{BHP}.
The answer to this fundamental question can be obtained in experiments on the search for neutrinoless double $\beta$-decay of some even-even nuclei.

\item
What is the value of the minimal neutrino mass ?

Neutrino oscillation experiments allow to determine only neutrino mass-squared differences.
Information on the minimal neutrino mass can be obtained from the future tritium KATRIN experiment and from cosmological data .
\item
What is the value of the parameter $\sin\theta_{13}$?

\item
What is the value of the leptonic CP phase?

The answer to this question apparently can be obtained in future experiments
at the JPARC facility \cite{JPARC}, in the off-axes neutrino experiments 
\cite{Feld}
in the $\beta$-beam neutrino experiments (see \cite{betabeam}) and in 
the Neutrino Factory experiments (see \cite{NF}).

\item
Is the number of massive neutrino equal to the number of flavor neutrinos (three)?

Indication in favor of $\bar\nu_{\mu} \to \bar\nu_{e}$ oscillations with
the third ``large'' neutrino mass-squared difference ($\Delta m^2\simeq 1 \rm{eV}^2$)
was obtained in the LSND experiment \cite{LSND}. If this indication  will be confirmed,
 we need (at least) four massive neutrinos to describe experimental data.
The situation apparently will be clarified by the MiniBooNE experiment
at Fermilab \cite{MiniB}.

\end{itemize}
It is my pleasure to thank  W. Alberico, M. Fabbrichesi and S.Petcov for useful discussions. I acknowledge the support of  the program ``Rientro dei cervelli''.


\begin{thebibliography}{99}

\bibitem{S-K} Super-Kamiokande Collaboration, S.~Fukuda {\it et al.,}
Phys. Rev. Lett. {\bf 81}, 1562 (1998);\,
 S.~Fukuda {\it et al.,} Phys. Rev. Lett. {\bf 82}, 2644 (1999);\,
S.~Fukuda {\it et al.,} Phys. Rev. Lett. {\bf 85}, 3999-4003 (2000).




\bibitem{SNO} SNO collaboration, Q.R. Ahmad {\it et al.}, Phys. Rev. Lett. 
 {\bf 87}, 071301 (2001)\,~
Q.R. Ahmad {\it et al.}, Phys.Rev.Lett. 
{\bf 89}, 011301 (2002); nucl-ex/0204008. \,~Q.R. Ahmad {\it et al.,} 
Phys.Rev.Lett 
{\bf 89}, 011302 (2002); nucl-ex/0204009.\,~Phys.Rev.Lett 
to be published, nucl-ex/0309004.


\bibitem{Kamland} KamLAND collaboration, K. Eguchi {\it et al.}, Phys. Rev. 
Lett. {\bf 90}, 021802 (2003) \,~ hep-ex/0212021.

\bibitem{Cl}B. T. Cleveland {\it et al.}, Astrophys. J. {\bf
496} (1998) 505.

\bibitem{GALLEX} GALLEX Collaboration, W. Hampel 
{\it et al.}, Phys. Lett. {\bf B 447} (1999) 127 ;\,
GNO Collaboration,
M. Altmann {\it et al.}, Phys. Lett. {\bf B 490} (2000) 16 ;\,
Nucl.Phys.Proc.Suppl. {\bf 91} (2001) 44.

\bibitem{SAGE} SAGE Collaboration,
J. N. Abdurashitov {\it et al.},
 Phys. Rev. {\bf C 60} (1999) 055801 ; \,Nucl.Phys.Proc.Suppl. {\bf 110}
(2002) 315;
\bibitem{S-Ksol} Super-Kamiokande Collaboration, S.~Fukuda {\it et al.}, Phys. Rev. Lett.
 {\bf 86} (2001) 5651;\, M.Smy, hep-ex/0208004.



\bibitem{Soudan} Soudan 2 Collaboration, W.W.M.Allison \textit{et al.}, 
Physics Letters {\bf B 449} (1999) 137; 

\bibitem{MACRO} MACRO Collaboration, M.Ambrosio et al.
hep-ex/0106049;\, Phys. Lett. B517 (2001) 59 \,~M. Ambrosio et al.
NATO Advanced Research Workshop on Cosmic Radiations,
Oujda (Morocco), 21-23 March, 2001.
\bibitem{Dirac1} P.A.M. Dirac, Proc. Roy. Soc. {\bf A 126} (1930) 369. 


\bibitem{WBook} S. Weinberg ``The Quantum Theory of Fields'' vol. I, Cambridge University Press, 2000.

\bibitem{Dirac3} P.A.M. Dirac, Proc. Roy. Soc. {\bf A 133} (1931) 60. 
 

\bibitem{Anderson} C.D. Anderson, Science {\bf 76}  (1932) 238.


\bibitem{Chadwick}. J.Chadwick, Proc. Roy. Soc.,{\bf  A,
136} (1932) 692.






\bibitem{Wu} C.S. Wu {\it et al.}, Phys. Rev.  {\bf 105} (1957) 1413.


\bibitem{LeeYang1} T.D. Lee and C.N. Yang,  Phys. Rev.  {\bf 104} (1956) 254. 

\bibitem{Landau}
L. Landau,
Nucl. Phys. \textbf{3} (1957) 127 (1957).

\bibitem{LeeYang2}
T.D. Lee and C.N. Yang,
Phys. Rev. \textbf{105} (1957) 1671 .

\bibitem{Salam}
A. Salam,
Il Nuovo Cim. \textbf{5} (1957) 299 .

\bibitem{Pauli} W.Pauli, Handbuch der Physik. vol.24, Part I (Springer, Berlin, 1933)







\bibitem{Goldhaber}
M. Goldhaber, L. Grodzins and A.W. Sunyar,
Phys. Rev. \textbf{109} (1958) 1015 .

\bibitem{FG}
R.P. Feynman and M. Gell-Mann,
Phys. Rev. \textbf{109} (1958) 193 .

\bibitem{Sudarshan-Marshak}
E.C.G. Sudarshan and R. Marshak,
Phys. Rev. \textbf{109} (1958)  1860 .

\bibitem{Danby} G. Danby  {\it et al.} Phys. Rev. Lett. \textbf{9} (1962) 36 .



\bibitem{BPontecorvo59} B. Pontecorvo, JETP \textbf{37} (1959) 1751 .



\bibitem{Cabibbo}
N.Cabibbo,
Phys. Rev. Lett. \textbf{10} (1963) 531.


\bibitem{GIM} S. L. Glashow, J. Iliopoulos and L. Maiani, Phys. Rev.  \textbf{D2} (1970) 1285.

\bibitem{BP78}
S.M. Bilenky and B.Pontecorvo,
Phys. Rep. \textbf{41} (1978) 225 .


\bibitem{Glashow61}
S.L. Glashow,
Nucl. Phys. \textbf{22} (1961) 597 .

\bibitem{Weinberg67}
S. Weinberg,
Phys. Rev. Lett. \textbf{19} (1967) 1264 .

\bibitem{Salam68}
A. Salam,
Proc. of the 8$^{\mathrm{th}}$ Nobel Symposium on \textit{Elementary Particle
  Theory, Relativistic Groups and Analyticity}, edited by N. Svartholm, 1969.

\bibitem{K-M}
M. Kobayashi and T. Maskawa,
Prog. Theor. Phys. \textbf{49} (1973) 652.

\bibitem{P57}
B.~Pontecorvo,
J. Exptl. Theoret. Phys. \textbf{33} (1957) 549.
[Sov. Phys. JETP \textbf{6} (1958) 429 ].

\bibitem{P58}
B.~Pontecorvo,
J. Exptl. Theoret. Phys. \textbf{34} (1958) 247 
[Sov. Phys. JETP \textbf{7} (1958) 172 ].



\bibitem{ReinesCowan} F.Reines and C.L. Cowan, Phys. Rev.  \textbf{92} (1953) 830;
Phys. Rev.  \textbf{113} (1959) 273 .

\bibitem{Davis}  R. Davis, Bull. Am. Phys. Soc., Washington meeting (1959)

\bibitem{P67}
B.~Pontecorvo,
Zh. Eksp. Teor. Fiz. \textbf{53} (1967) 1717 
[Sov. Phys. JETP \textbf{26} (1968) 984].



\bibitem{Giunti} C. Giunti, hep- ph/0311241.

\bibitem{BPet}
S.M. Bilenky and S.T. Petcov,
Rev. Mod. Phys. \textbf{59} (1987) 671.

\bibitem{BilGiunti} S. M. Bilenky and C.  Giunti,
 Int. J. Mod. Phys. \textbf{A16} (2001) 3931, hep-ph/0102320.


\bibitem{Mainz} Ch. Weinheimer , Proceedings of 
the 20th International Conference on Neutrino
Physics and Astrophysics, {\em Neutrino 2002}\,(Munich, Germany)
May 25-30, 2002.



\bibitem{Troitsk} V. Lobashev \textit{et al.}, Nucl. Phys. Proc. Suppl. 
{\bf 91 }(2001) 280.

\bibitem{Sloan} M. Tegmark et al., astro-ph/0310723.

\bibitem{K2K} K2K collaboration, M.H.Ahn {\it et al.}, Phys.Rev.Lett. 90 (2003) 041801; 
hep-ex/0212007. 

\bibitem{SSM} J. N. Bahcall, M. H. Pinsonneault and S. Basu,
Astrophys. J. {\bf 555} (2001). 990 .

\bibitem{MSW} L. Wolfenstein, Phys. Rev. D \textbf{17} (1978) 2369 ; 
Phys. Rev. D \textbf{20} (1979) 2634;
S.P. Mikheyev and A.Yu. Smirnov, Yad. Fiz. \textbf{42} (1985) 1441  [Sov. J. Nucl. Phys. \textbf{42} (1985) 913 ]; Il Nuovo Cim. C \textbf{9} (1986) 17 ; Zh. Eksp. Teor. Fiz. \textbf{91} (1986) 7  [Sov. Phys. JETP \textbf{64}(1986) 4].

\bibitem{MNS}
Z. Maki, M. Nakagawa and S. Sakata,
Prog. Theor. Phys. \textbf{28} (1962) 870.



\bibitem{CHOOZ} CHOOZ Collaboration,  M.\, Apollonio \textit{et al.}, 
Phys. Lett. B {\bf 466} (1999) 415.

\bibitem{PVerde} F. Boehm, J. Busenitz et al., Phys.\ Rev.\ Lett.\  {\bf 84} (2000) 3764 ;
\, Phys. Rev. D {\bf 62} (2000) 072002.
 
\bibitem{Schramm} X. Shi and D.N. Schramm, Phys. Lett.  \textbf{ B 283} (1992) 305.


\bibitem{BGiu} S.M. Bilenky  and C.Giunti,
Phys. Lett.\textbf{ B444} (1998) 379-386;
hep-ph/9802201.




\bibitem{13} http://www.hep.anl.gov/minos/reactor13/white.html


\bibitem{MINOS} MINOS Collaboration, K. Lang \textit{et al.}, 
Int.J.Mod.Phys.\textbf{ A18} (2003 ) 3857.


\bibitem{JPARC} T. Kobayashi, Nucl.Phys.Proc.Suppl. 111 (2002) 163

\bibitem{Doi} M. Doi, T. Kotani and E. Takasugi, Progr. Theor. Phys. Suppl.
 \textbf{53} (1985) 1.

\bibitem{BPet} S.M. Bilenky  and S.T. Petcov,
Rev. Mod. Phys.\textbf{59} (1987) 671.

\bibitem{Elliott} S. R. Elliott and P. Vogel, Annu. Rev. Nucl. Part.
 Sci. \textbf{52} (2002), hep-ph/0202264.

\bibitem{BGGM}
S.M. Bilenky, C. Giunti, J.A. Grifols and E. Masso,  Phys. Rept. \textbf{379} (2003) 69-148, hep-ph/0211462.

\bibitem{PP} S. Pascoli , S.T. Petcov,
Proceedings of 10th International Workshop on Neutrino Telescopes, Venice, Italy, 11-14 Mar 2003 (vol. 1p. 301);\,~
 hep-ph/0308034.




\bibitem{Gratta} G. Gratta, Proceedings of the XXI International
Symposium on Lepton and Photon Interactions at High Energies, 1-16 August 2003,
Fermilab, Batavia, Illinois USA.


\bibitem{H-M} Heidelberg-Moscow collaboration, H. V. Klapdor-Kleingrothaus
\textit{et al.}, Eur. Phys. J. \textbf{ A 12},(2001) 147.

\bibitem{Katrin} KATRIN collaboration, A.Osipowicz \textit{et al.}, 
hep-ex/0109033.


\bibitem{see-saw}
M. Gell-Mann, P. Ramond and R. Slansky,
in \textit{Supergravity}, p.~315, edited by F. van Nieuwenhuizen and D.
  Freedman, North Holland, Amsterdam, 1979;\,~
T.~Yanagida,
Proc. of the \textit{Workshop on Unified Theory and the Baryon Number of the
  Universe}, KEK, Japan, 1979;\,~
R.N. Mohapatra and G.~Senjanovi{\'c},
Phys. Rev. Lett. \textbf{44}, 912 (1980).

\bibitem{Buch}W. Buchmuller, P. Di Bari, M. Plumacher, hep-ph/0401240.


\bibitem{BHP}S.M. Bilenky, J. Hosek and S.T. Petcov, 
Phys. Lett.\textbf{B94} (1980) 495. 

\bibitem{Feld} G. Feldman , Proceedings of 
the International Workshop ``Neutrino oscillations in Venice'',' 3-5 December, 2003.





\bibitem{betabeam} J. Burguet-Castell, D. Casper, J.J. Gomez-Cadenas, P.Hernandez, F.Sanchez, 
hep-ph/0312068.



\bibitem{NF} M. Lindner 
Int.J.Mod.Phys.\textbf{A18} (2003) 3921.



\bibitem{LSND} LSND Collaboration,
A. Aguilar  \textit{et al.}, Phys.Rev.\textbf{D64} (2001) 112007;\,hep-ex/0104





\bibitem{MiniB}
MiniBooNE Collaboration W.C. Louis \textit{et al.}, 
Proceeding of 10th International Workshop on Neutrino Telescopes, 
Venice, Italy, 11-14 Mar 2003 (vol. I, p. 181).













\end{thebibliography}
\end{document}